\documentclass[twocolumn,prl,showpacs,superscriptaddress]{revtex4}%
\usepackage{amssymb}
\usepackage{amsmath}
\usepackage{amsfonts}
\usepackage{graphicx}%
\begin{document}
\title{Deterministic and Efficient Quantum Cryptography Based on Bell's Theorem}
\author{Zeng-Bing Chen}
\affiliation{Hefei National Laboratory for Physical Sciences at Microscale and Department
of Modern Physics, University of Science and Technology of China, Hefei, Anhui
230026, China}
\affiliation{Physikalisches Institut, Universit\"{a}t Heidelberg, Philosophenweg 12, 69120
Heidelberg, Germany}
\author{Qiang Zhang}
\affiliation{Hefei National Laboratory for Physical Sciences at Microscale and Department
of Modern Physics, University of Science and Technology of China, Hefei, Anhui
230026, China}
\author{Xiao-Hui Bao}
\affiliation{Hefei National Laboratory for Physical Sciences at Microscale and Department
of Modern Physics, University of Science and Technology of China, Hefei, Anhui
230026, China}
\author{J\"{o}rg Schmiedmayer}
\affiliation{Physikalisches Institut, Universit\"{a}t Heidelberg, Philosophenweg 12, 69120
Heidelberg, Germany}
\author{Jian-Wei Pan}
\affiliation{Hefei National Laboratory for Physical Sciences at Microscale and Department
of Modern Physics, University of Science and Technology of China, Hefei, Anhui
230026, China}
\affiliation{Physikalisches Institut, Universit\"{a}t Heidelberg, Philosophenweg 12, 69120
Heidelberg, Germany}


\begin{abstract}
We propose a novel double-entanglement-based quantum cryptography protocol
that is both efficient and deterministic. The proposal uses photon pairs with
entanglement both in polarization and in time degrees of freedom; each
measurement in which both of the two communicating parties register a photon
can establish one and only one perfect correlation and thus deterministically
create a key bit. Eavesdropping can be detected by violation of local realism.
A variation of the protocol shows a higher security, similarly to the
six-state protocol, under individual attacks. Our scheme allows a robust
implementation under current technology.

\end{abstract}
\date{}
\maketitle

Entanglement and nonlocality lie at the heart of modern understanding of
quantum foundations. One of the most striking aspects of entanglement is that
certain \textit{statistical} correlations derived for entangled states can be
in conflict with local realism \cite{EPR}, as quantitatively shown by Bell's
inequalities (BI) \cite{Bell}. These fundamental issues were originally
considered at the very boundary of physics and philosophy. Yet, they have
found practical applications in quantum information science. In a remarkable
paper by Ekert \cite{Ekert}, BI have a profound utility in quantum
cryptography (QC) (or, quantum key distribution, QKD)
\cite{BB84,BBM,B92,Gisin-RMP}. Actually, there is a fascinating link
\cite{Ekert,Gisin-RMP,Bell-QC,Bell-Gisin-PRL} between security of certain
quantum communication protocols and BI. However, quantum violations of local
realism also occur in an \textquotedblleft
all-versus-nothing\textquotedblright\ (AVN)\ form
\cite{GHZ-90,Pan-GHZ,Cabello,Chen}, which is more striking in the sense that
the contradiction between quantum mechanics and local realism arises even for
\textit{definite }correlations. It thus remains to be seen if such an AVN
nonlocality can have any application in quantum information, particularly, in
QC. Most of the QC protocols \cite{BB84,BBM,B92,Gisin-RMP} proposed so far are
non-deterministic as only less than 50\% qubits detected can be further used
as key bits. This may be a practical problem, e.g., in the one-time-pad
secret-key cryptosystem \cite{Gisin-RMP}. Such a problem may be eliminated by
deterministic QC protocol and secret direct communication, which attracted
some recent interest \cite{cryp-pp,Zhao}.

For QC experiments realized with faint laser pulses, they may be insecure
under the so-called beamsplitter (BS) attack \cite{Gisin-RMP}. This is because
the currently available photon sources have a finite probability of emittting
more than one photon (or more than one entangled photon pair for entangled
photon sources). An eavesdropper (usually called Eve) could in principle use a
channel with lower photon loss or without loss and only allow those attenuated
pulses containing $n$ ($n\geq2$) photons to reach the receiver. For these
pulses she can use a BS to steal at least one photon, thus getting full
information without being detected. However, the entanglement-based QC
protocols (such as Ekert's \cite{Ekert} and ours to be described below)
exploit entanglement as certain \textquotedblleft security
resource\textquotedblright\ and do not suffer from this kind of problem as the
security therein is guaranteed by the violations of BI.

In this paper, based on the previously proved two-party AVN nonlocality (or
inseparability) for two doubly-entangled photon pairs \cite{Chen}, we propose
a novel QC protocol which is efficient and deterministic: \textit{Each
detected photon pair can establish a key bit} with the help of classical
communications. This deterministic feature of our protocol stems from the very
nature of the two-party AVN nonlocality: The two communicating parties always
have perfect quantum correlations for whatever measurement bases they choose.
An eavesdropper can be detected by observing the violation of local realism
for the quantum channel. A variation of the present protocol is similar to the
six-state protocol \cite{six-state} and shows a higher security under simple
individual attacks. A remarkable advantage of the present scheme is that all
required measurements can be done with linear optical elements and as such,
the experimental realization of the protocol is within the reach of current
technology.%
\begin{figure}
[ptb]
\begin{center}
\includegraphics[
height=2.0821in,
width=2.6646in
]%
{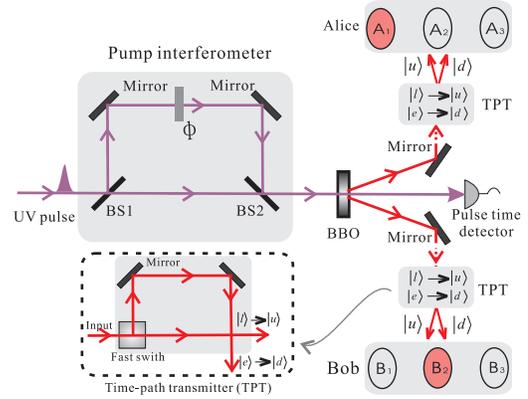}%
\caption{A proposed implementation of our QKD protocol.}%
\end{center}
\end{figure}

In our protocol (see Fig. 1) \textquotedblleft
doubly-entangled\textquotedblright\ photon pairs (photon-$1$ and photon-$2$)
in the state
\begin{equation}
\left\vert \Psi\right\rangle _{12}=\frac{1}{2}(\left\vert H\right\rangle
_{1}\left\vert H\right\rangle _{2}+\left\vert V\right\rangle _{1}\left\vert
V\right\rangle _{2})(\left\vert \uparrow\right\rangle _{1}\left\vert
\uparrow\right\rangle _{2}+\left\vert \downarrow\right\rangle _{1}\left\vert
\downarrow\right\rangle _{2})\label{source}%
\end{equation}
are generated via spontaneous parametric down conversion (SPDC) and sent,
respectively, to two communicators, Alice and Bob. Here $\left\vert
H\right\rangle $ ($\left\vert V\right\rangle $) stands for photons with
horizontal (vertical) polarization; $\left\vert \uparrow\right\rangle $ and
$\left\vert \downarrow\right\rangle $ span an orthonormal basis for either
time or path states of photons. $\left\vert \Psi\right\rangle _{12}$ is
maximally entangled both in polarization and in time/path degrees of freedom
of photons. The creation of polarization-path entanglement was discussed in
\cite{Chen,Simon}. With a pump interferometer in Fig. 1 one can generate
polarization-time double entanglement (More details are given at the end of
this paper).

To create secure keys, each party needs to measure observables involving the
spin-type operators $x=\left\vert H\right\rangle \left\langle V\right\vert
+\left\vert V\right\rangle \left\langle H\right\vert $, $z=\left\vert
H\right\rangle \left\langle H\right\vert -\left\vert V\right\rangle
\left\langle V\right\vert $ (for polarization) and $x^{\prime}=\left\vert
\uparrow\right\rangle \left\langle \downarrow\right\vert +\left\vert
\downarrow\right\rangle \left\langle \uparrow\right\vert $, $z^{\prime
}=\left\vert \uparrow\right\rangle \left\langle \uparrow\right\vert
-\left\vert \downarrow\right\rangle \left\langle \downarrow\right\vert $ (for
time/path). Particularly, the two parties should measure nine observables:
\begin{equation}%
\begin{array}
[c]{ccc}%
\begin{array}
[c]{c}%
A_{1}\left\{
\begin{array}
[c]{c}%
z_{1}\\
x_{1}^{\prime}\\
z_{1}\cdot x_{1}^{\prime}%
\end{array}
,\right. \\
\text{ \ }%
\end{array}
&
\begin{array}
[c]{c}%
A_{2}\left\{
\begin{array}
[c]{c}%
z_{1}^{\prime}\\
x_{1}\\
x_{1}\cdot z_{1}^{\prime}%
\end{array}
,\right. \\
\text{ \ }%
\end{array}
&
\begin{array}
[c]{c}%
A_{3}\left\{
\begin{array}
[c]{c}%
z_{1}z_{1}^{\prime}\\
x_{1}x_{1}^{\prime}\\
z_{1}z_{1}^{\prime}\cdot x_{1}x_{1}^{\prime}%
\end{array}
;\right. \\
\ \
\end{array}
\\
B_{1}\left\{
\begin{array}
[c]{c}%
x_{2}^{\prime}\\
x_{2}\\
x_{2}\cdot x_{2}^{\prime}%
\end{array}
,\right.  & B_{2}\left\{
\begin{array}
[c]{c}%
z_{2}\\
z_{2}^{\prime}\\
z_{2}\cdot z_{2}^{\prime}%
\end{array}
,\right.  & B_{3}\left\{
\begin{array}
[c]{c}%
z_{2}x_{2}^{\prime}\\
x_{2}z_{2}^{\prime}\\
z_{2}x_{2}^{\prime}\cdot x_{2}z_{2}^{\prime}%
\end{array}
.\right.
\end{array}
\label{aa}%
\end{equation}
Here Alice (Bob) arranges her (his) local observables into three groups
$A_{1}$, $A_{2}$\ and $A_{3}$ ($B_{1}$, $B_{2}$\ and $B_{3}$), each of which
has three operators. As in Ref. \cite{Chen}, the three operators of each group
can be measured by one and the same apparatus (to be described below). This is
crucial in the AVN argument of nonlocality without the necessity of an
additional assumption of noncontextuality \cite{Chen,Lvovsky}. When measuring
the three operators of each group, e.g., $A_{1}$ (other groups are similar),
one measures $z_{1}$ and $x_{1}^{\prime}$ simultaneously with the apparatus
(also labeled as $A_{1}$), thus also giving the measurement result of
$z_{1}\cdot x_{1}^{\prime}$, which is just the product of the readouts of
$z_{1}$ and $x_{1}^{\prime}$. To denote this fact we then use $(\cdot)$ to
separate operators (as in $z_{1}\cdot x_{1}^{\prime}$, $x_{1}\cdot
z_{1}^{\prime}$, $x_{2}\cdot x_{2}^{\prime}$ and $z_{2}\cdot z_{2}^{\prime}$)
or operator products (as in $z_{1}z_{1}^{\prime}\cdot x_{1}x_{1}^{\prime}%
$\ and $z_{2}x_{2}^{\prime}\cdot x_{2}z_{2}^{\prime}$) that can be identified
as local \textquotedblleft elements of reality\textquotedblright\ in the
nonlocality argument \cite{Chen}. In this way,\ the three operators in each
group\ are co-measurable and measured simultaneously by the same apparatus.
Totally, one thus requires six apparatuses ($A_{1}$, $A_{2}$\ and $A_{3}$ for
Alice; $B_{1}$, $B_{2}$\ and $B_{3}$ for Bob), which can be realized without
any mutual conflict only by linear optical elements \cite{Chen}.

Now we are ready to describe the present QC protocol. For each of the emitted
pairs, photon-1 (photon-2) goes to Alice (Bob) who then measures an operator
group, which is chosen \textit{randomly and independently} from the three
groups $A_{1}$, $A_{2}$\ and $A_{3}$ ($B_{1}$, $B_{2}$\ and $B_{3}$). Any
local outcome of the above measurements is completely random and can of course
be either $-1$\ or $+1$, representing thus one bit of information.

Now one immediately has the following: For each pair of operator groups chosen
by Alice and Bob, there is one and only one pair of outcomes of the local
operators (or operator products) that possesses perfect correlation; totally
Alice and Bob can establish nine pairs of perfectly correlated local outcomes
as each of the two parties has three operator groups. For instance, if Alice
(Bob) measures the three operators in $A_{1}$ ($B_{2}$), then only the
outcomes of $z_{1}$ and $z_{2}$\ will show perfect correlation, i.e., their
product will certainly be $1$. The above result stems from the fact that for
the photon pairs in $\left\vert \Psi\right\rangle _{12}$ one has the following
nine eigenequations \cite{Chen}
\begin{align}
&  \left.  z_{1}\cdot z_{2}\left\vert _{\left\vert \Psi\right\rangle _{12}%
}\right.  =1,\;z_{1}^{\prime}\cdot z_{2}^{\prime}\left\vert _{\left\vert
\Psi\right\rangle _{12}}\right.  =1,\right. \label{e1}\\
&  \left.  x_{1}\cdot x_{2}\left\vert _{\left\vert \Psi\right\rangle _{12}%
}\right.  =1,\;x_{1}^{\prime}\cdot x_{2}^{\prime}\left\vert _{\left\vert
\Psi\right\rangle _{12}}\right.  =1,\right. \label{e2}\\
&  \left.  z_{1}z_{1}^{\prime}\cdot z_{2}\cdot z_{2}^{\prime}\left\vert
_{\left\vert \Psi\right\rangle _{12}}\right.  =1,\ \ x_{1}x_{1}^{\prime}\cdot
x_{2}\cdot x_{2}^{\prime}\left\vert _{\left\vert \Psi\right\rangle _{12}%
}\right.  =1,\right. \label{e3}\\
&  \left.  z_{1}\cdot x_{1}^{\prime}\cdot z_{2}x_{2}^{\prime}\left\vert
_{\left\vert \Psi\right\rangle _{12}}\right.  =1,\ \ x_{1}\cdot z_{1}^{\prime
}\cdot x_{2}z_{2}^{\prime}\left\vert _{\left\vert \Psi\right\rangle _{12}%
}\right.  =1,\right. \label{e4}\\
&  \left.  z_{1}z_{1}^{\prime}\cdot x_{1}x_{1}^{\prime}\cdot z_{2}%
x_{2}^{\prime}\cdot x_{2}z_{2}^{\prime}\left\vert _{\left\vert \Psi
\right\rangle _{12}}\right.  =-1.\right.  \label{e5}%
\end{align}
Here we have used a simplification in notions, e.g., $z_{1}\cdot
z_{2}\left\vert _{\left\vert \Psi\right\rangle _{12}}\right.  =1$ means
$z_{1}\cdot z_{2}\left\vert \Psi\right\rangle _{12}=\left\vert \Psi
\right\rangle _{12}$.

After the above measurements have taken place on a photon pair, Alice and Bob
can announce in public by classical communications which of the three operator
group they have measured. They discard all measurements in which either or
both of them fail to register a photon at all. In the case where Alice and Bob
have detected a photon simultaneously from the emitted photon pairs, they can
establish \textit{deterministically} a secure key as they can know from the
classical communications which pair of their outcomes has perfect correlation.
For example, let us again assume that Alice (Bob) has chosen the apparatus
$A_{1}$ ($B_{2}$). In this case the two parties will certainly obtain
$z_{1}\cdot z_{2}\left\vert _{\left\vert \Psi\right\rangle _{12}}\right.  =1$,
from which Alice using her own outcome of $z_{1}$ can predict with certainty
Bob's outcome of $z_{2}$, and \textit{vise versa}. Any one of this type of
perfect correlations can then be used to create deterministically a secure key
bit. The deterministic feature of our QC protocol is thus demonstrated. As a
comparison, Ekert's protocol is nondeterministic in the sense that successful
detection of a photon by both Alice and Bob can establish at most $2/9$ raw key.

All QKD protocols are consisted of two parts: the quantum part producing the
raw keys and the classical part (e.g., reconciliation and privacy
amplification) \cite{Gisin-RMP}. The later is not considered here as it is the
same for all cryptographic protocols \cite{high-D}. Now it is ready to see
that our protocol is more effective than traditional protocols (e.g., Ekert's
protocol) in the quantum part. For comparison, in Ekert's protocol $7/9$ of
detected photon pairs is of no use for establishing raw keys and will be
sacrificed to detect eavesdropping. Thus, in the quantum part our protocol is
$1/(1-7/9)=4.5$ times more efficient than the original Ekert protocol.

A complete security analysis of our QKD protocol is very difficult and beyond
the scope of this paper. Here we consider the security issue by first
following Ekert's security analysis \cite{Ekert}. In Ekert's protocol, the
presence of an eavesdropper can be detected in conjunction with BI. This is
because a possible intervention (interception, detection, and substitution of
photons) by the eavesdropper is equivalent to introducing local elements of
physical reality into the system. Following this line of thought, Eve's
intervention would acquire information, e.g., by randomly measuring
observables like $A\in\{A_{1},A_{2},A_{3}\}$ and $B\in\{B_{1},B_{2},B_{3}\}$
with certain results (denoted by $\lambda$); afterwards she sends the
replacement of the detected photons to Alice and Bob. Now what Alice and Bob
do is just to measure certain predetermined values of these operators (i.e.,
elements of physical reality) as measured already by Eve. In this case, Alice
and Bob would obtain the correlation $E_{A\cdot B}=\int d\lambda\rho
(\lambda)E_{A}(\lambda)E_{B}(\lambda)$, where the integration may also be
summation if the number of $\lambda$ is finite and $\rho(\lambda)$ is the
probability for Eve's result $\lambda$. Thus, in the presence of Eve, one has
\cite{Cabello,Chen}
\begin{align}
\left\langle \mathcal{O}\right\rangle _{Eve} &  \equiv\int d\lambda
\rho(\lambda)\left[  -E_{z_{1}z_{1}^{\prime}\cdot x_{1}x_{1}^{\prime}\cdot
z_{2}x_{2}^{\prime}\cdot x_{2}z_{2}^{\prime}}+E_{z_{1}z_{1}^{\prime}\cdot
z_{2}\cdot z_{2}^{\prime}}\right.  \nonumber\\
&  +E_{x_{1}x_{1}^{\prime}\cdot x_{2}\cdot x_{2}^{\prime}}+E_{z_{1}\cdot
x_{1}^{\prime}\cdot z_{2}x_{2}^{\prime}}+E_{x_{1}\cdot z_{1}^{\prime}\cdot
x_{2}z_{2}^{\prime}}\nonumber\\
&  \left.  \left.  +E_{z_{1}\cdot z_{2}}+E_{z_{1}^{\prime}\cdot z_{2}^{\prime
}}+E_{x_{1}\cdot x_{2}}+E_{x_{1}^{\prime}\cdot x_{2}^{\prime}}\right]
\leq7.\right.  \label{oeve}%
\end{align}
Here $E_{A\cdot B}=[C(A\cdot B=+1)-C(A\cdot B=-1)]/[C(A\cdot B=+1)+C(A\cdot
B=-1)]$, where $C(A\cdot B=\pm1)$ are the counting numbers when the measured
variable $A\cdot B=\pm1$. The measured nine sets of perfect correlations allow
Alice and Bob to infer $E_{A\cdot B}$ in (\ref{oeve}) and, then, $\left\langle
\mathcal{O}\right\rangle _{Eve}$. The observed violation of (\ref{oeve}) can
thus detect Eve's eavesdropping by randomly choosing a small portion
(\textquotedblleft fair sample\textquotedblright) of the generated key bits.
Note that quantum prediction of the upper bound of (\ref{oeve}) can be $9$
\cite{Cabello,Chen}, as can be seen from Eqs. (\ref{e1})-(\ref{e5}).

The present scheme for detecting the eavesdropper is conceptually striking in
the following sense. In Ekert's protocol, perfect correlations are used to
establish secure keys, while statistical correlations are used to detect
eavesdropping in terms of BI. However, in our protocol \textit{perfect
correlations play the dual role of both establishing secure keys and detecting
eavesdroppers}. We have thus demonstrated the link for the security against
eavesdropping of our protocol and a two-party version of Bell's theorem, and
the definite quantum predictions used in a two-party AVN nonlocality argument
may have fascinating application in the deterministic QKD protocol.

Note that $\left\vert \Psi\right\rangle _{12}$ is a maximally entangled state
in a $4\otimes4$ dimensional Hilbert space \cite{Chen}. To achieve higher
security in QKD protocols, one may use either high-dimensional systems
\cite{high-D} or more alternative settings (e.g., three-base protocol
\cite{six-state}). Thus one might expect that our QKD protocol using three
measurement bases per party and high-dimensional entanglement has a bonus of
higher security. To show that this is indeed the case, recall that Ekert's
protocol can be regarded as a variation \cite{BBM} of the BB84 protocol
\cite{BB84}, and as such the security of the former can be guaranteed by the
security of the latter. Similarly, let us consider the case where Alice
prepares the doubly-entangled pair herself, measures one of her three operator
groups in (\ref{aa}), and sends to Bob photon-2, which might be subject to
Eve's intercept-resend attacks. This modified protocol is then, in some sense,
similar to the six-state (three-base) protocol \cite{six-state}, which is more
secure than the origianl BB84 protocol. For instance, when Alice measures
$A_{1}$, she will collapse her state onto the basis vectors $\left\vert
H\right\rangle _{1}\left\vert \bar{\uparrow}\right\rangle _{1}$ ($z_{1}%
=1,x_{1}^{\prime}=1$) or $\left\vert V\right\rangle _{1}\left\vert
\bar{\downarrow}\right\rangle _{1}$ ($z_{1}=-1,x_{1}^{\prime}=-1$) for which
$z_{1}\cdot x_{1}^{\prime}=1$, or $\left\vert H\right\rangle _{1}\left\vert
\bar{\downarrow}\right\rangle _{1}$ ($z_{1}=1,x_{1}^{\prime}=-1$) or
$\left\vert V\right\rangle _{1}\left\vert \bar{\uparrow}\right\rangle _{1}$
($z_{1}=-1,x_{1}^{\prime}=1$) for which $z_{1}\cdot x_{1}^{\prime}=-1$. Here
$\left\vert \bar{\uparrow}\right\rangle =\frac{1}{\sqrt{2}}(\left\vert
\uparrow\right\rangle +\left\vert \downarrow\right\rangle )$ and $\left\vert
\bar{\downarrow}\right\rangle =\frac{1}{\sqrt{2}}(\left\vert \uparrow
\right\rangle -\left\vert \downarrow\right\rangle )$. If Alice gets
$\left\vert H\right\rangle _{1}\left\vert \bar{\uparrow}\right\rangle _{1}$
(with the probability of $1/4$), Bob's state will be equivalently prepared as
$\left\vert H\right\rangle _{2}\left\vert \bar{\uparrow}\right\rangle _{2}$,
which is exactly the equal-amplitude superposition of the basis vectors for
any of $\{B_{1},B_{2},B_{3}\}$. Note that any two basis vectors $\left\vert
e_{\alpha}\right\rangle $ and $\left\vert e_{\beta}\right\rangle $ belonging
to different bases in $\{B_{1},B_{2},B_{3}\}$ satisfy $\left\vert \left\langle
e_{\beta}\right.  \left\vert e_{\alpha}\right\rangle \right\vert ^{2}=1/4$,
i.e., the three bases $\{B_{1},B_{2},B_{3}\}$ are mutually unbiased. If Eve,
with the probability of $2/3$, uses wrong bases, she gets wrong perfect
correlations with Alice and thus no information. Explicit calculation shows
that Eve can be detected with the probability of $1/2$ in this case. Thus
Bob's error rate under the simple individual attacks is $1/3$, implying that
our protocol might be more secure, similarly to the six-state protocol, but
eliminates the latter's disadvantage of low efficiency.

A recent experiment \cite{Yang} (see also \cite{PanPurification}) has
successfully created the path-polarization-entangled\ two-photon states. Note
that the two photons experiences two different paths from the source to the
detectors. Then the coherence of the path entanglement will be sensitive to
the relative phase that a photons would acquire as it propagates along the two
paths. The unavoidable fluctuations in the relative phase may destroy the path
entanglement. To maintain the path coherence, especially in the long-distance
case, the long-distance interferometric stability is required, which is
extremely difficult in practice.

Fortunately, one can overcome the above problem by using pulsed entanglement
source where the two photons are entangled both in time (i.e., time-bin
entanglement \cite{time,time-50km}) and in polarization. To create the
required entanglement, a short, ultraviolet (UV) laser pulse is sent first
through an unbalanced Mach-Zehnder interferometer (the pump interferometer)
and then through a BBO crystal (see Fig. 1). The pump pulse is splitted by the
first (50\%-50\%) BS (BS1) into two pulses, one propagating along the short
path and another along the long path. If the pulse duration is shorter than
the arm length difference, the output from (50\%-50\%) BS2 is two pulses well
separated in time. For the case where there is one and only one
polarization-entangled pair [assumed to be in $\frac{1}{\sqrt{2}}(\left\vert
H\right\rangle _{1}\left\vert H\right\rangle _{2}+\left\vert V\right\rangle
_{1}\left\vert V\right\rangle _{2})$ for definiteness] production after the
\textquotedblleft early\textquotedblright\ and \textquotedblleft
late\textquotedblright\ pulses pass through the BBO crystal, the
polarization-time entangled\ two-photon state $\left\vert \Psi\right\rangle
_{12}=\frac{1}{2}(\left\vert H\right\rangle _{1}\left\vert H\right\rangle
_{2}+\left\vert V\right\rangle _{1}\left\vert V\right\rangle _{2})(\left\vert
e\right\rangle _{1}\left\vert e\right\rangle _{2}+\left\vert l\right\rangle
_{1}\left\vert l\right\rangle _{2})$ is then created by adjusting the phase
$\phi$. Here $\left\vert \uparrow\right\rangle \equiv\left\vert e\right\rangle
$ (early time) and $\left\vert \downarrow\right\rangle \equiv\left\vert
l\right\rangle $ (late time) are two orthonormal time states of photons. In
Fig. 1 the pulse time detector can determine the emission time of the pump
laser, giving a time fiducial signal.

Now each photon hold by Alice or Bob propagates along the same path. In this
way, the time-entanglement is much more robust than the path-entanglement.
Indeed, time-bin entanglement has been experimentally distributed over 50 km
in optical fibers \cite{time-50km}. However, time-bin measurement is
non-deterministic \cite{time,time-50km} and may thus reduce the efficiency of
the key production.

Here we propose a measurement scheme with simple linear optical elements and
fast switches. The setup in Fig. 1 can measure all local observables in
(\ref{aa}) by using two \textquotedblleft time-path
transmitters\textquotedblright\ (TPT) with optical paths identical to the pump
interferometer. In the TPT a fast swith will reflect an incident photon into
the long path of the TPT only for photons in $\left\vert e\right\rangle $;
otherwise, it is swithed off so that the $\left\vert l\right\rangle $ photons
simply propagate along the short path of the TPT. The fast swith is controlled
according to the timing of the pulsed photons by noting that $\left\vert
e\right\rangle $ and $\left\vert l\right\rangle $ are two time states
distinguishable with respect to the time fiducial signal (see Fig. 1). In this
way, the TPT transforms \textit{coherently} $\left\vert e\right\rangle $
($\left\vert l\right\rangle $) to $\left\vert d\right\rangle $ ($\left\vert
u\right\rangle $), with $\left\vert d\right\rangle $ and $\left\vert
u\right\rangle $ being two distinguishable paths of photons. Afterwards, all
measurements in (\ref{aa}) can be done by the linear optics setups in Ref.
\cite{Chen}.

The function of a fast switch can be accomplished by an acousto-optic
modulator (AOM). Due to bulk acousto-optic interaction, an incident laser beam
can be either diffracted (\textquotedblleft first order\textquotedblright) by
or directly transmitted (\textquotedblleft zero order\textquotedblright)
through an acousto-optic medium, depending on whether the acoustic wave is
present or not. Thus, if $\left\vert e\right\rangle $ ($\left\vert
l\right\rangle $) is subject to the first-order (zero-order) process,
$\left\vert e\right\rangle $ and $\left\vert l\right\rangle $ will be
separated in path, acting exactly as a TPT. The intensity change (the
wavelength change can be safely neglected) between the zero-order and
first-order beams may be compensated by an attenuator. The current commercial
AOM \cite{AOM} can reach a rising time of about several nanoseconds, which is
sufficient enough for our proposal. Moreover, it was already used as a fast
optic switch (On/Off), e.g., by Kuzmich \textit{et al}. \cite{AOM} for
generating nonclassical photon pairs.

To summarize, we have proposed a double-entanglement-based QC protocol that is
both efficient and deterministic. The deterministic feature and high
efficiency of our protocol have obvious advantages in a practical utility.
Importantly, our protocol is within the reach of current technology and even
allows a robust intermediate-distance realization.

\textit{Note added in proof}.---Recently, M. Genovese kindly informed us their
related work (Ref. 25) using path-time double entanglement in nondeterministic QKD.

We thank Y.-A. Chen, X.-B. Wang and Z. Zhao for useful discussions. This work
was supported by the Fok Ying Tung Education Foundation, the NNSFC and the
CAS. We also acknowledge funding by the European Commission under Contract No.
509487 and the Alexander von Humboldt Stiftung.

\end{document}